\documentclass[pra,aps,showpacs,twocolumn,nofootinbib,floatfix]{revtex4}
\usepackage{hyperref}
\usepackage{amssymb}
\usepackage{amsbsy}
\usepackage{graphicx}
\usepackage{amsmath}
\usepackage{times,txfonts}

\newcommand{\tr}{\text{tr}}
\newcommand{\gr}[1]{\boldsymbol{#1}}
\newcommand{\sig}{\boldsymbol{\sigma}}

\begin{document}
\title{Gaussian interferometric power}

\author{Gerardo Adesso}
\affiliation{$\mbox{School of Mathematical Sciences, The University of
Nottingham, University Park, Nottingham NG7 2RD, United Kingdom}$}

\date{August 1, 2014}

\begin{abstract}
The interferometric power of a bipartite quantum state quantifies the precision, measured by quantum Fisher information, that such a state enables for the estimation of a parameter embedded in a unitary dynamics applied to one subsystem only, in the worst-case scenario where a full knowledge of the generator of the dynamics is not available {\it a priori}. For finite-dimensional systems, this quantity was proven to be a faithful measure of quantum correlations beyond entanglement. Here we extend the notion of interferometric power to the technologically relevant setting of optical interferometry with continuous-variable probes. By restricting to Gaussian local dynamics, we obtain a closed formula for the interferometric power of all two-mode Gaussian states. We identify separable and entangled Gaussian states which maximize the interferometric power at fixed mean photon number of the probes, and discuss the associated metrological scaling. At fixed  entanglement of the probes, highly thermalized states can guarantee considerably larger precision than pure two-mode squeezed states.
\end{abstract}

\maketitle
\section{Introduction}

The second quantum revolution \cite{secondquantum} is dawning. The long-anticipated fundamental advantages brought about by quantum technologies in applications such as secure communication, precise sensing and metrology, are starting to materialize thanks mainly to the impressive progresses in the experimental control of light and matter at the quantum level \cite{nobelsummary}. On the other hand, some very central issues are still to be addressed at the theoretical level, which can be summarized into one straight question: Which quantum features are ultimately needed to outperform the operation of classical devices?

Metrology \cite{helstrom} is one field where considerable debate around such a question has been spurned in recent years.  While in some metrological setups entangled probes can lead to an extra gain in precision for the estimation of unobservable parameters compared to separable probes \cite{giovannetti}, such an enhancement can fade away under the most common sources of noise \cite{susana,opticmal,escher,madalin}. At the same time, other means to achieve supraclassical performances even without using entanglement have been devised \cite{boixo,napolitano}.  Somehow disappointingly, one might then conclude that quantum correlations in the form of entanglement are neither necessary nor sufficient for quantum-enhanced metrology in general.

Here we focus on a specific,  highly relevant metrological setting, namely optical interferometry \cite{caves}. The most pressing mission of optical interferometry is arguably the revelation of weak phase shifts induced by gravitational waves \cite{ligo,ligonew,ligolimits}. Optical interferometric setups traditionally involve a Mach--Zender interferometer, in which a relative phase is acquired between the two arms and needs to be detected at the output \cite{caves}. It is convenient to model theoretically such a setup as a dual-arm channel, where a phase shift is applied to one arm only, while the identity operation is applied to the other arm \cite{opticmal}. Note that, in practice, the implementation of the scheme requires an additional phase reference beam; see e.g.~the discussion in \cite{noterefidiota}. If, as it is customary, the generator of the phase shift is known {\it a priori}, then tailored nonclassical resources such as single-mode squeezed states or two-mode entangled states can be exploited to improve the precision of phase estimation beyond the classical shot noise level. The mathematical techniques for assessing the ultimate precision limits allowed by quantum mechanics for parameter estimation are beautifully rooted in information geometry and find widespread applications \cite{wootters,quantumcramer,paris}.

\begin{figure}[t!]
\includegraphics[width=8.5cm]{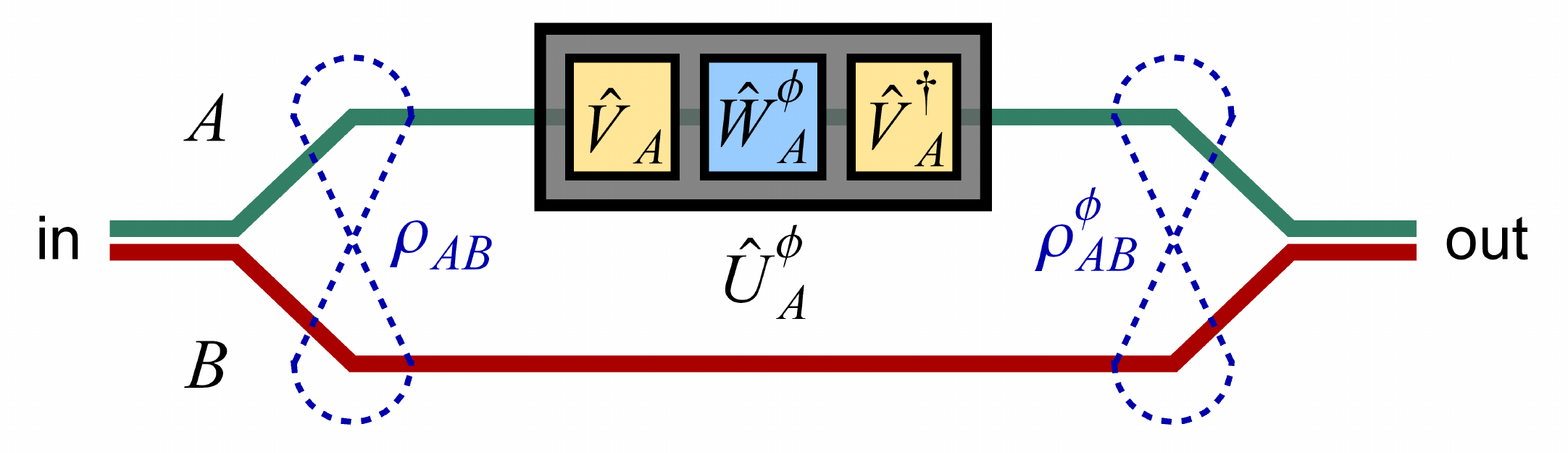}
\caption{(Color online) Black box optical interferometry.}
\label{gipfig}
\end{figure}

In this paper, we explore optical interferometry in a black box setting where the generator of the phase shift on one arm is not known {\it a priori}.  The aim of this analysis is to elucidate exactly which characteristics of continuous-variable quantum states are necessary and sufficient for them to act as sensitive probes to not just one, but a variety of possible local dynamics. We identify such essential characteristics with genuinely quantum correlations between the two modes entering the interferometer, of a general type commonly referred to as quantum discord \cite{oz,hv,adessodatta,giordaparis,modirev}. This finding resonates with the analogous one for finite-dimensional systems, where the novel paradigm of black box metrology has been very recently introduced \cite{ip}.

The worst-case precision enabled by a two-party probe state in a black box phase estimation setting defines the interferometric power of the state (Section~\ref{sec2}). Remarkably, we obtain a closed formula for this operational quantifier in the relevant instance of two-mode Gaussian probes (which include squeezed and thermal states), thus assessing their potential usefulness for practical sensing technologies (Section~\ref{sec3}). We then identify Gaussian states which offer, in principle, optimized performances in optical metrology with or without entanglement and even in the presence of high thermal noise at the probe preparation stage (Section~\ref{sec4}).

\section{Black box optical interferometry} \label{sec2}

Let us begin by formalizing the black box paradigm \cite{ip} for optical interferometry \cite{caves,opticmal}.
We consider a bosonic continuous-variable system of two modes $A$ and $B$, respectively described by the annihilation operators $\hat{a}$ and $\hat{b}$. We can define a vector of quadrature operators (in natural units, $\hbar=1$) as $\hat{{\gr{O}}}=\{\hat{q}_A, \hat{p}_A, \hat{q}_B, \hat{p}_B\}$, where $\hat{q}_A=(\hat{a}+ \hat{a}^\dagger)/{\sqrt2}$ and $\hat{p}_A=(\hat{a}- \hat{a}^\dagger)/{i\sqrt2}$ (and similarly for mode $B$). The canonical commutation relations are then compactly expressed as $[\hat{O}_j,\hat{O}_k] = i \Omega_{kl}$, with $\gr \Omega={{\ 0\ \ 1}\choose{-1\ 0}}^{\oplus 2}$ being the two-mode symplectic form  \cite{ourreview}.

A two-mode state $\rho_{AB}$ is prepared as the input of an interferometer, see Fig.~\ref{gipfig}. Mode $A$ enters a black box where it undergoes a unitary transformation
\begin{equation}\hat{U}^\phi_{A} = \exp(i \phi \hat{H}_A)\,,
\end{equation} whose full specifics are unknown {\it a priori}.
In analogy with the finite-dimensional case \cite{ip}, we need to restrict the generator $\hat{H}_A$ to have a nondegenerate spectrum, in order to avoid trivial dynamics. In the present continuous-variable setting, the most natural and maximally informative choice for the spectrum of $\hat{H}_A$ is a harmonic one. With this prescription, we can then decompose the black box transformation as follows without any loss of generality,
\begin{equation}\label{blackbox}
\hat{U}^\phi_A = \hat{V}^\dagger_A \hat{W}^\phi_A \hat{V}_A\,,
\end{equation}
where $\hat{W}^\phi_A= \exp(i \phi \hat{n}_A)$ is a standard phase shift generated by the number operator $\hat{n}_A = \hat{a}^\dagger \hat{a}$, and $\hat{V}_A$ is an arbitrary unitary transformation.
The transformed state of the two modes after the black box is
\begin{equation}\label{rhophi}
\rho^{\phi,\hat{V}_A}_{AB} = (\hat{U}^\phi_A \otimes \mathbb{I}_B)  \rho_{AB} (\hat{U}^\phi_A \otimes \mathbb{I}_B)^{\dagger}\,.
\end{equation}
One can then perform a measurement on the output state, to construct an estimator $\phi_{\rm est}$ for the parameter $\phi$. One can iterate the probing trial a large number $\kappa$ of times (or equivalently, one can run $\kappa$ parallel experiments if one has the availability of $\kappa$ independent copies of the black box), to improve the statistical accuracy of the estimator. In mathematical terms, the variance of any estimator for the parameter $\phi$, defined as $\Delta\phi^2 \equiv \langle ({\phi}_{\rm est} - \phi)^2\rangle$, is constrained by the Cram\'er-Rao bound \cite{quantumcramer},
\begin{equation}\label{cramerrao}
\kappa \Delta\phi^2 \geq \frac{1}{{\cal F}(\rho^\phi_{AB})}\,,
\end{equation}
where the quantity at the denominator is known as the quantum Fisher information (QFI) \cite{quantumcramer,paris} and can be interpreted as the squared speed of evolution of the probe state $\rho_{AB}^\phi$ under an infinitesimal transformation $(\hat{U}^{\epsilon}_A \otimes \mathbb{I}_B)$ \cite{wootters,petz}. Under a smoothness hypothesis, the QFI can be defined as \cite{mariona,pinel}
\begin{equation}\label{qfifid}
{\cal F}(\rho^\phi_{AB}) = -2 \lim_{\epsilon \rightarrow 0} \frac{\partial^2 F(\rho_{AB}^\phi, \rho_{AB}^{\phi+\epsilon})}{\partial \epsilon^2}\,,
\end{equation}
via the Uhlmann fidelity  \cite{uhlmann}
\begin{equation}
\label{ulmafid}
F(\rho_1, \rho_2) = \left\{\tr\big[({\sqrt{\rho_1} \rho_2 \sqrt{\rho_1}})^\frac12\big]\right\}^2\,.
\end{equation}
For single-parameter estimation, the bound in (\ref{cramerrao}) is asymptotically tight (for $\kappa \gg 1$). This means that the QFI directly quantifies the precision (intended as the inverse of the variance of the estimator per trial) that can be achieved with the input probe state $\rho_{AB}$, for the estimation of the parameter $\phi$ embedded in the local transformation $\hat{U}^\phi_{A}$, by means of a specific optimized measurement on the output state $\rho_{AB}^\phi$. For this reason, the QFI is conventionally adopted as the figure of merit in quantum metrology \cite{paris}.

With this in mind,
the interferometric power (IP) of the state $\rho_{AB}$, with respect to the probing mode $A$, is then defined as
\begin{equation}\label{ipcv}
{\cal P}^A(\rho_{AB}) = \frac14 \inf_{\hat{V}_A} {\cal F}(\rho^{\phi,\hat{V}_A}_{AB})\,,
\end{equation}
where the $\frac14$ is a normalization factor adopted here for consistency with the finite-dimensional definition of IP \cite{ip}.
The quantity ${\cal P}^A(\rho_{AB})$ evaluates the worst-case precision guaranteed  by using $\rho_{AB}$ as a probe, where the minimization runs over all possible choices of local dynamics generated by a Hamiltonian $\hat{H}_A$ with harmonic spectrum. In practice, probe states $\rho_{AB}$ with higher IP embody more reliable resources for metrology, as they ensure a smaller variance in the estimation of $\phi$ even if uncontrollable unitary fluctuations   $\hat{V}_A$ occur in conjunction with the designed phase shift $\hat{W}_A^\phi$; in general, this can happen even in absence of entanglement \cite{modix,lqu,ip}.

Notice that, by definition, the IP is invariant under local unitary operations, ${\cal P}^A[(\hat{U}_A' \otimes \hat{U}_B'')\rho_{AB} (\hat{U}_A' \otimes \hat{U}_B'')^\dagger] = {\cal P}^A[\rho_{AB}]$ \cite{ip}. This follows by observing that unitaries on $B$ are irrelevant for the QFI, while unitaries on $A$ can be reabsorbed in the minimization of Eq.~(\ref{ipcv}). Notice however that, in spite of the convexity of the QFI, the IP is {\it not} convex. One can namely show the following inconclusive chain of inequalities. Given two states $\rho_{AB}$ and $\tau_{AB}$ and a probability $0 \leq p \leq 1$, one has
\begin{eqnarray*}
&&4 {\cal P}^A[p \rho_{AB} + (1-p) \tau_{AB}] \\
& \leq & {\cal F}[p \rho^{\phi,\hat{V}_A}_{AB} + (1-p) \tau^{\phi,\hat{V}_A}_{AB}] \\
& \leq & p {\cal F}[\rho^{\phi,\hat{V}_A}_{AB}] + (1-p) {\cal F}[\tau^{\phi,\hat{V}_A}_{AB}] \\
& \geq & 4 p {\cal P}^A[\rho_{AB}] + 4(1-p) {\cal P}^A[\tau_{AB}]\,,
\end{eqnarray*}
where we used the definition of IP in the first and last inequalities, and the convexity of the QFI in the middle one.
In particular, one can construct straightforward examples where a state with nonzero IP is obtained by mixing two states $\rho_{AB}$ and $\tau_{AB}$ with zero IP; this happens when the minimum $\hat{V}_A$ is different for $\rho_{AB}$ and $\tau_{AB}$.

\section{Gaussian IP: Definition and properties}\label{sec3}

In the following, we restrict our attention to a fully Gaussian scenario. Namely, the probe states $\rho_{AB}$ are assumed to be two-mode Gaussian states, and the local dynamics $\hat{U}_{A}^\phi$ is assumed to be Gaussian (also known as linear), i.e., preserving the Gaussian character of the states it acts upon. It is in order to recall that a Gaussian state
$\rho_{AB}$ is represented by a Gaussian characteristic function in phase space, and is completely specified by the first and second moments of the quadrature operators \cite{ourreview}, collected respectively in the vector $\gr{\delta}_{AB} = ({\delta}_j)$ and in the covariance matrix  $\sig_{AB} = ({\sigma}_{jk})$, where $\delta_j = \tr[\rho_{AB} \hat{O}_j]$ and $\sigma_{jk} = \tr[\rho_{AB}\{(\hat{O}_j-\delta_j),  (\hat{O}_k -\delta_k)\}_+]$ (with $j,k=1,\ldots,4$). As the first moments can be adjusted by local displacements, and since the IP is invariant under local unitary operations, in what follows we can consider without any loss of generality states with zero first moments $\gr{\delta}=\gr{0}$, described entirely by their covariance matrices. The latter will correspond to physical states in the Hilbert space provided the {\it bona fide} condition
\begin{equation}\label{bonafide}
\sig_{AB} + i {\gr \Omega} \geq 0\,,
 \end{equation}
 which incarnates the Robertson-Schr\"odinger uncertainty relation, is satisfied.

Concerning the local dynamics, the Gaussianity restriction amounts to requiring that the generator $\hat{H}_A$ be at most quadratic in the canonical operators $\hat{a}, \hat{a}^\dagger$. Given the decomposition in (\ref{blackbox}), and noting that $\hat{W}_A^\phi$ is already a Gaussian unitary, this requirement is passed on $\hat{V}_A$. In general, up to irrelevant displacements, a Gaussian unitary $\hat{V}_A$ is associated via the metaplectic representation to a symplectic transformation (i.e.~a real matrix which preserves the symplectic form) acting by congruence on covariance matrices \cite{pramana}. By virtue of this correspondence, together with well established results of symplectic algebra and Gaussian quantum information \cite{ourreview,pirlareview}, we can now translate the scheme of Fig.~\ref{gipfig} and the above equations at the phase space level, as follows.

The probe state $\rho_{AB}$ will be described by its covariance matrix $\sig_{AB}$. The black box unitary $\hat{U}_{A}^\phi$ corresponds to a symplectic transformation ${\gr T}_A^\phi={\gr M}^{\sf T}_A {\gr R}^\phi_A {\gr M}_A$, with
\begin{equation}\label{rot}
{\gr R}_A^\phi = \left(
\begin{array}{cc}
 \cos (\phi ) & -\sin (\phi ) \\
 \sin (\phi ) & \cos (\phi ) \\
\end{array}
\right)
\end{equation}
being a phase-space rotation of an angle $\phi$ in phase space.
Furthermore, ${\gr M}_A$ can be written in general according to the Euler decomposition \cite{pramana,serafozzinazi}, ${\gr M}_A = {\gr R}^\psi_A {\gr S}^\zeta_A {\gr R}^\theta_A$, where ${\gr S}^\zeta_A = \text{diag}(\zeta, 1/\zeta)$, with $\zeta > 0$, is a squeezing transformation. In this way, Eq.~(\ref{blackbox}) translates into
${\gr T}_A^{\phi, \zeta,\theta} = {{\gr R}^\theta_A}^{\sf T} {\gr S}^\zeta_A {\gr R}^\phi_A {\gr S}^\zeta_A {\gr R}^\theta_A$.
From Eq.~(\ref{rhophi}), the transformed state after the black box has a covariance matrix given by
\begin{equation}\label{sigmaoutsi}
\sig_{AB}^{\phi,\zeta,\theta} = ({\gr T}_A^{\phi, \zeta,\theta} \oplus {\gr I}_B) \sig_{AB} ({\gr T}_A^{\phi, \zeta,\theta} \oplus {\gr I}_B)^{\sf T}\,.
\end{equation}
The Gaussian IP is of a two-mode Gaussian probe with covariance matrix $\sig_{AB}$ is then defined as
\begin{equation}\label{ipg}
{\cal P}_G^A(\sig_{AB}) = \frac14 \inf_{\zeta, \theta} {\cal F}(\sig_{AB}^{\phi,\zeta,\theta})\,.
\end{equation}

The fidelity $F$ between two (undisplaced) two-mode Gaussian states, Eq.~(\ref{ulmafid}), which enters in the definition (\ref{qfifid}) of the QFI, can be computed from the respective covariance matrices $\sig_1$ and $\sig_2$ as \cite{marian}
\begin{equation}\label{marianfid}
F(\sig_1,\sig_2)=\big\{\sqrt{\Gamma
   }+\sqrt{\Lambda }-[(\sqrt{\Gamma }+\sqrt{\Lambda })^2-\Upsilon ]^\frac12\big\}^{-1}\,,
   \end{equation}
where
\begin{eqnarray*}
\Gamma&=&16 \det[{\gr \Omega} (\sig_1/2) {\gr \Omega} (\sig_2/2)],\\
\Lambda &=&16 \det[(\sig_1+i {\gr \Omega})/2] \det[(\sig_2+i {\gr \Omega})/2],\\
\Upsilon &=& \det[(\sig_1+\sig_2)/2].
\end{eqnarray*} Notice that alternative yet related studies of QFI for Gaussian states can be found e.g.~in Refs.~\cite{mariona,pinel,alex}.

We now recall that, by local symplectic operations, every two-mode covariance matrix
\begin{equation}\label{cm}
\sig_{AB}=\left(\begin{array}{cc}
\gr\alpha & \gr\gamma \\
\gr\gamma^{\sf T} & \gr\beta
\end{array}\right),
\end{equation}
can be transformed in a standard form with all diagonal $2 \times 2$ subblocks,  $\gr\alpha=\text{diag}(a,a)$, $\gr\beta=\text{diag}(b,b)$, $\gr\gamma=\text{diag}(c,d)$,
where $a,b \geq 1, c \geq |d| \geq 0$.
Exploiting once more the invariance of the (Gaussian) IP under local unitaries, we now proceed to evaluate Eq.~(\ref{ipg}) on probe states with covariance matrix in standard form. In such case, the minimization over $\theta$ in (\ref{ipg}) turns out to be solved simply by $\theta=0$. The value of $\zeta$ which yields the minimum in (\ref{ipg}) is instead less trivial, and can be written as an analytical yet too cumbersome function of $a,b,c,d$ to be reported here \cite{notecunt}.

After some tedious algebra, we arrive at one of the main results of this paper: a closed formula for the Gaussian IP of all two-mode Gaussian states. This is independent of the standard form used for the explicit evaluation, and can be recast in terms of the four local symplectic invariants of an arbitrary covariance matrix, defined as $A=\det\gr\alpha$, $B=\det\gr\beta$, $C=\det\gr\gamma$, and $D=\det\sig_{AB}$. The formula reads
\begin{equation}\label{ipgg}
{\cal P}^A_G(\sig_{AB})=\frac{X+\sqrt{X^2+Y Z}}{2Y}\,,
\end{equation}
where
\begin{eqnarray*}
X&=&(A+C)(1+B+C-D)-D^2\,, \\
Y&=&(D-1)(1+A+B+2C+D)\,, \\
Z&=&(A+D)(AB-D)+C(2A+C)(1+B)\,.
\end{eqnarray*}

We can now analyze the properties of the Gaussian IP for two-mode Gaussian states. In \cite{ip}, the IP has been proven to capture a peculiar nonclassical feature of bipartite states of a finite-dimensional system: their amount of quantum correlations beyond entanglement, of the so-called discord type \cite{modirev}. We will now show that the same interpretation holds in the infinite-dimensional Gaussian case. First of all, the Gaussian IP vanishes if and only the state is a zero-discord state (also known as classical-quantum state) \cite{modirev}. In the Gaussian case, under a natural constraint of bounded mean energy per mode, the only classical-quantum states are product states \cite{adessodatta,giordaparis,ralph}. From Eq.~(\ref{ipgg}), we find indeed that the only two-mode Gaussian states with vanishing Gaussian IP are product states, characterized by the invariants $C=0, D=AB$. All correlated two-mode Gaussian states are therefore useful for black box optical interferometry, returning a nonzero QFI for any possible local Gaussian dynamics. Furthermore, the Gaussian IP is invariant under local unitary operations as already mentioned, and it can be shown to be monotonically nonincreasing under arbitrary Gaussian quantum operations on subsystem $B$. The proof follows from the definition of QFI and can be adapted from the finite-dimensional case \cite{ip}. Namely, suppose a Gaussian probe state with covariance matrix $\sig_{AB}'$ is obtained from the state with covariance matrix $\sig_{AB}$ by the action of a completely positive trace-preserving and Gaussianity-preserving map (a Gaussian quantum channel) on $B$. Any such a map commutes with the unitary phase shift applied on $A$, so it can be moved after the black box and considered as part of the output measurement. Since  ${\cal F}(\sig_{AB}^{\phi,\zeta,\theta})$ defines the optimal precision achieved by the best possible output measurement, the Fisher information associated to $\sig_{AB}'^{\phi,\zeta,\theta}$ can only be smaller or equal, which proves the claim.

Altogether, these properties allow us to conclude that the Gaussian IP is a faithful measure of bipartite discord-type correlations \cite{modirev} for Gaussian states. With the result of Eq.~(\ref{ipgg}), the IP becomes the only currently known faithful measure of discord-type correlations which is computable both for two-qubit states \cite{ip} and for two-mode Gaussian states, which are respectively the pillars of discrete-variable and continuous-variable bipartite quantum information processing.

  Eq.~(\ref{ipgg}) acquires a very simple form for states characterized by $d=\mp c$ in standard form (in which case the optimal $\zeta$ in (\ref{ipg}) reduces to $1$), which include the relevant classes of squeezed thermal states ($d=-c$) and mixed thermal states ($d=c$):
\begin{equation}\label{ipgsts}
{\cal P}^A_G(\left.\sig_{AB}\right\vert_{d=\mp c})=\frac{c^2}{2(a b-c^2 \pm 1)}\,.
\end{equation}
Notice that in this simple case the Gaussian IP is symmetric under swapping of the two modes $A$ and $B$, but this is not true for general two-mode Gaussian states, as clear from Eq.~(\ref{ipgg}). For pure states, specified by $b=a, -d=c=\sqrt{a^2-1}$, one has in particular
${\cal P}^A_G(\left.\sig_{AB}\right\vert_{b=a, -d=c=\sqrt{a^2-1}})=(a^2-1)/4$, which is a monotonic function of the marginal mixedness of each subsystem. This  means that the Gaussian IP reduces to a Gaussian entanglement monotone \cite{ourreview} on pure states. This is, once more, a desired property for a discord-type quantifier, and holds analogously in the finite-dimensional case \cite{ip}.

\begin{figure}[t!]
\includegraphics[height=5.3cm]{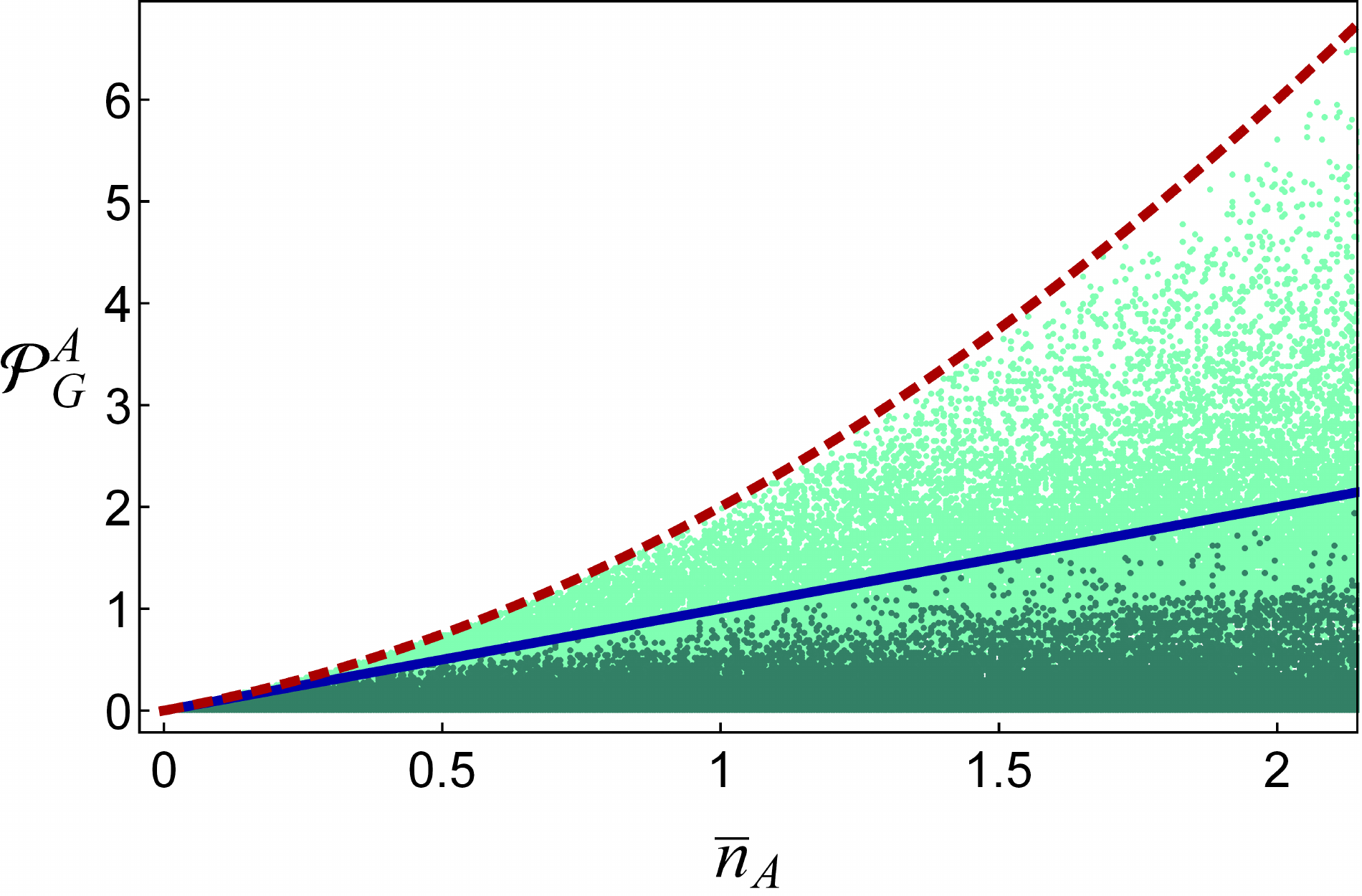}
\caption{(Color online) Gaussian IP versus  mean photon number of the probing mode $A$ for $10^5$ entangled (lighter) and separable (darker) Gaussian states. The standard quantum limit ${\cal P}_G^A = \bar{n}_A$ (solid line) and the Heisenberg limit ${\cal P}_G^A = \bar{n}_A (\bar{n}_A+1)$ (dashed line) are indicated. }
\label{gipvsn}
\end{figure}

\section{Gaussian IP versus local mean photon number and entanglement}\label{sec4}

It is particularly interesting to study the scaling of the worst-case QFI, namely the Gaussian IP, with the mean photon number of the probing subsystem $A$, \begin{equation}\label{nbarra}
{\bar n}_A \equiv \tr[\rho_{AB} (\hat{n}_A \otimes \mathbb{I}_B)]=\frac{\tr\, {\gr \alpha}-2}{4}\,,\end{equation}
 which conventionally defines the resource count in optical interferometry \cite{caves,giovannetti,opticmal}. A numerical exploration of random two-mode Gaussian states $\sig_{AB}$, as shown in Fig.~\ref{gipvsn}, reveals two distinct regimes. As expected, separable probe states can never surpass the standard quantum limit (or shot noise limit), given by a linear scaling of the IP with $\bar{n}_A$; entangled states, on the other hand, can have IP scaling at most quadratically with $\bar{n}_A$, reaching up to the so-called Heisenberg limit.
A class of states with the maximum possible Gaussian IP in absence of entanglement, for instance, is given in standard form by $d=c=\sqrt{(a-1)(b-1)}$, in the limit $b \gg 1$; for these states, $$\lim_{b \rightarrow \infty}
{\cal P}^A_G(\left.\sig_{AB}\right\vert_{d=c=\sqrt{(a-1)(b-1)}}) = {\bar n}_A\,,$$ spanning the solid
line in Fig.~\ref{gipvsn}. Entangled states with maximum Gaussian IP at fixed $\bar{n}_A$ are instead pure two-mode squeezed states,  sitting on the dashed line in Fig.~\ref{gipvsn}, for which (as mentioned before) $${\cal P}^A_G(\left.\sig_{AB}\right\vert_{b=a, -d=c=\sqrt{a^2-1}}) = {\bar n}_A({\bar n}_A+1)\,.$$ However, there are a considerable number of entangled states which perform worse than shot noise, which means that their entanglement does not translate into a practical quantum enhancement for black box metrology.

\begin{figure}[t!]
\includegraphics[height=5.3cm]{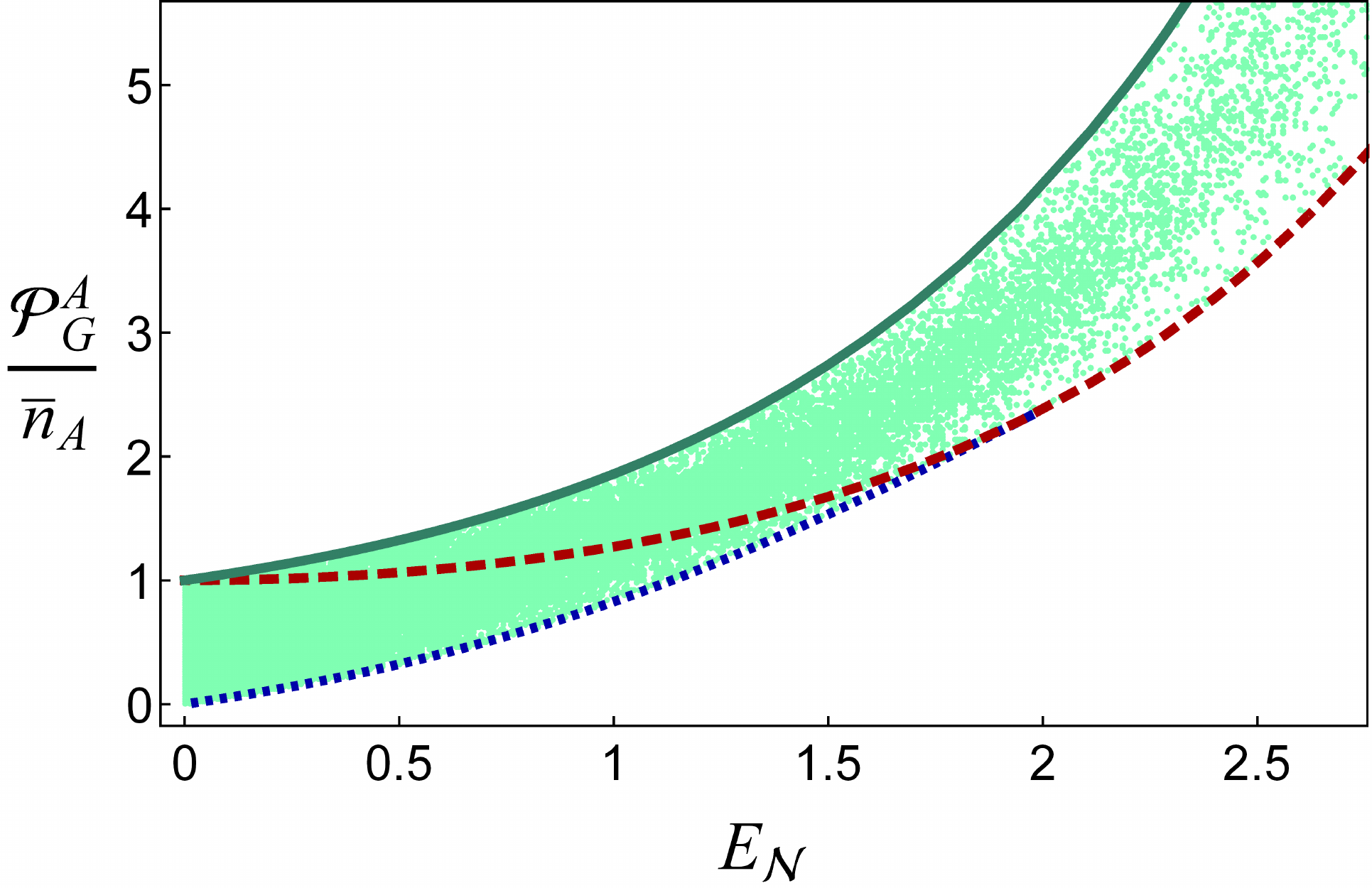}
\caption{(Color online) Gaussian IP normalized by the mean photon number of mode $A$, plotted versus the logaritmic negativity $E_{\cal N}$ for $10^5$ entangled Gaussian states. The dashed line accommodates pure states. See text for details of the other boundary curves.}
\label{gipvse}
\end{figure}

Motivated by the above observation, we perform a thorough analysis of the interplay between the Gaussian IP ${\cal P}^A_G$, rescaled by the local mean photon number $\bar{n}_A$, and the entanglement of two-mode Gaussian states. The latter can be conveniently measured by the logarithmic negativity \cite{vidalwerner,pleniorompe}, which is a decreasing function of the smallest symplectic eigenvalue $\tilde{\nu}$ of the partial transpose of the covariance matrix,
\begin{equation}\label{lone}
E_{\cal N}(\sig_{AB}) = \max\{0,\, -\ln \tilde{\nu}\}\,,
\end{equation}
where $2\tilde{\nu}^2 = H - \sqrt{H^2-4D}$ with $H = A+B-2C$ \cite{ourreview}.
Fig.~\ref{gipvse} shows a comparison between the two quantities, which reveals that
${\cal P}^A_G/\bar{n}_A$ is bounded from above and from below at fixed $E_{\cal N}$.
To derive the expression of the bounds analytically, we start from the ansatz that the extremal states are to be found within the class of entangled squeezed thermal states. We can reparametrize their standard form covariances as $-d=c=\sqrt{(a-\tilde\nu)(b-\tilde\nu)}$, with $0<\tilde\nu<1$, and perform a constrained optimization of $${\cal P}^A_G/\bar{n}_A=[ (a-\tilde\nu ) (b-\tilde\nu )]/[(a-1) (a \tilde\nu +b \tilde\nu -\tilde\nu ^2+1)]$$ at fixed $\tilde\nu$, subject to the  {\it bona fide} condition (\ref{bonafide}).

We then find that the upper boundary (solid line) in Fig.~\ref{gipvse} is given by states with $a=(1+b-b\tilde\nu+\tilde\nu^2)/(1+\tilde\nu)$ in the limit $b \rightarrow \infty$, for which $$({\cal P}^A_G/\bar{n}_A)^{\rm up}\rightarrow (1+\tilde{\nu})/(2\tilde\nu)\,.$$ The lower boundary has two branches, see Fig.~\ref{gipvse}. For $\tilde\nu_0 <\tilde \nu <1$ (where $\tilde \nu_0 \approx 0.14$ is the real root of the polynomial $\tilde{\nu}^3+\tilde{\nu}^2+7 \tilde{\nu}-1$), i.e.~$E_{\cal N} \lessapprox 2$, the extremal states (dotted line) have $a=[\sqrt{2(\tilde\nu +1)^{3}}+3 \tilde\nu +1]/(1-\tilde\nu), b=\sqrt{2(\tilde\nu+1)}+\tilde\nu+2$, for which  $$({\cal P}^A_G/\bar{n}_A)^{{\rm low}_1} = \left(\frac{2}{\tilde\nu +1}-\frac{2}{\tilde\nu -1}-\frac{2 \sqrt{2}}{\sqrt{\tilde\nu +1}}-1\right)^{-1}\,.$$ For $0<\tilde\nu<\tilde\nu_0$, i.e.~$E_{\cal N} \gtrapprox 2$, the extremal states (dashed line) are pure two-mode squeezed states, described by $a=b=(1+\tilde\nu^2)/2, -d=c=(1-\tilde\nu^2)/2$, for which
$$({\cal P}^A_G/\bar{n}_A)^{{\rm low}_2}
= \frac{(\tilde\nu +1)^2}{4 \tilde\nu}\,.$$

This analysis reveals several interesting facts which can be relevant for applications. First, there is a minimum threshold in entanglement to beat necessarily the shot noise limit in black box metrology: all two-mode Gaussian states with $E_{\cal N} \gtrapprox 1.135$ achieve ${\cal P}^A_G > \bar{n}_A$, while some less entangled states can be outperformed by separable, more discordant states. Second, pure states eventually offer the {\it worst} possible metrological performance in black box optical interferometry for a given (sufficiently high) degree of entanglement. Conversely, highly thermalized states such as the ones on the upper boundary of Fig.~\ref{gipvse} can attain significantly higher Gaussian IP per local mean photon number, at equal degree of entanglement.  This is a very practical situation where the combined effect of entanglement and state mixedness surprisingly results in an enhancement of discord-type correlations useful for an operative task (namely interferometry in this case), somehow giving shape to the abstract statistical predictions of Ref.~\cite{allnon}.

Finally, we like to point out that Fig.~\ref{gipvse} is comparable to Fig.~1 (right panel) of \cite{adessodatta}, which features the Gaussian entropic discord versus the Gaussian entanglement of formation, although the extremal states are different. In particular, for separable states both the Gaussian discord and the Gaussian IP divided by $\bar{n}_A$ can reach at most one \cite{adessodatta,giordaparis}, while they are unbounded for entangled states. Overall this confirms the intimate connection between IP and discord.

\section{Conclusions}

In conclusion, we extended the paradigm of black box parameter estimation to the technologically important setting of optical interferometry. We defined the operative notion of interferometric power for a two-mode probe system, and specialized its definition to the case of Gaussian states and local Gaussian phase dynamics. We derived a closed formula for the Gaussian interferometric power of all two-mode Gaussian states. By studying  it against the mean photon number and  the entanglement of the probes, we singled out classes of extremal Gaussian states which guarantee the best possible metrological precision in a worst-case scenario. These states can be highly thermalized, which eases the demands for their implementation in laboratory.

This work develops a conceptual and practical advance for the characterization and exploitation of general nonclassical correlations in continuous-variable systems, and complementing Ref.~\cite{ip} it shows that their role in metrology transcends specific schemes and Hilbert space dimensions. The formalism applied here can be immediately useful to calculate other discord-type quantities for Gaussian states, which capture geometrically their sensitivity under local unitary transformations, e.g.~the local quantum uncertainty \cite{lqu}, the discriminating strength \cite{distrength}, and the discord of response \cite{disresp} (see also \cite{stellaretc}).

{\it Note added.}---Upon completion of the present manuscript, a preprint appeared \cite{merda} where a similar measure is independently defined, and explicitly computed only for the subclass of symmetric two-mode squeezed thermal states.

\acknowledgments{Discussions with Antonella De Pasquale, Alessandro Farace, Vittorio Giovannetti, Davide Girolami, Fabrizio Illuminati, Luca Rigovacca, and Tommaso Tufarelli are warmly acknowledged. This work was supported by the Foundational Questions Institute (FQXi-RFP3-1317) and the Brazilian CAPES (Pesquisador Visitante Especial-Grant~No. 108/2012).}



\end{document}